# A new experiment to test parity symmetry in cold chiral molecules using vibrational spectroscopy

A. Cournol[1], M. Manceau[1], M. Pierens[1], L. Lecordier[1], D. B. A. Tran[1,†], R. Santagata[1,§], B. Argence[1,#], A. Goncharov[1,£], O. Lopez[1], M. Abgrall[2], Y. Le Coq[2], R. Le Targat[2], H. Álvarez Martinez[2,$], W. K. Lee[2], D. Xu[2], P-E Pottie[2], R. J. Hendricks[3,!], T. E. Wall[3], J. M. Bieniewska[3], B. E. Sauer[3], M. R. Tarbutt[3], A. Amy-Klein[1], S. K. Tokunaga[1], and B. Darquié[1]

**Abstract:**

We present a brief review of our progress towards measuring parity violation in heavy-metal chiral complexes using mid-infrared Ramsey interferometry. We discuss our progress addressing the main challenges, including the development of buffer-gas sources of slow, cold polyatomic molecules, and the frequency-stabilisation of quantum cascade lasers calibrated using primary frequency standards. We report investigations on achiral test species of which promising chiral derivatives have been synthesized.

1) **Introduction**

Many molecules are chiral, coming in left- and right-handed versions known as enantiomers. Enantiomers are mirror images of one another. The potential energy surface of a chiral molecule shows two minima separated by an interconversion barrier, which one may associate with the left- and right-handed enantiomers. Strikingly, enantiomers, or chiral states, are not the eigenstates of the symmetric molecular electromagnetic Hamiltonian and nutation from one to the other can happen by quantum tunnelling through the potential barrier. However, interaction with the environment, or decoherence, tends to hold the molecule in the left- or right-handed enantiomer. When the interconversion barrier between enantiomers is very high, which is the case for the molecules considered here, the left- and right-handed states can then be considered to a good approximation as energy eigenstates. In addition to the electromagnetic interaction, the weak interaction, one of the three other fundamental forces, is also at play in molecular systems, notably in the interaction between electrons and nuclei. According to the Standard Model, the energy levels of enantiomers should be slightly different because of the parity-violation (PV, the left-right symmetry breaking) inherent in the weak force [1]. We aim to make the first ever measurement of this symmetry-breaking energy difference. This measurement will serve as a sensitive probe of the weak force and the limits of the Standard Model [2], and might shed light on the mystery of biological homochirality [3]. After the first evidence in atomic experiments carried out in Novosibirsk that neutral weak currents violate parity [4], people have dreamt for decades of measuring the PV energy difference between chiral enantiomers, but no experiment has ever reached the required sensitivity. A number of techniques have been proposed for the observation of PV in chiral molecules, including rotational, rovibrational, electronic, Mössbauer and NMR spectroscopy, as well as crystallization and solubility experiments, or optical activity measurements (see [2,5-8] and references therein). However, to our knowledge very few other groups currently pursue an experimental realisation. Dmitry Budker's group has recently proposed a measurement using NMR spectroscopy [9] and Martin Quack *et al.* at ETH Zürich are currently pursuing a different approach to ours, based on the measurement of the time evolution of parity in chiral species after a parity selection step [10], with quite distinct experimental challenges and which requires working with substantially different chiral species. This group has published a proof of principle using an achiral molecule [11] and has also proposed promising candidate species for this experiment [12].

The parity-violating energy difference in turn leads to frequency differences between the rovibrational spectra of left- and right-handed enantiomers, which we aim to measure for the first time using precise mid-infrared spectroscopy [13]. The most precise experiment, carried out around 2000 at the Laboratoire de Physique des Lasers by one of the co-authors, measured the 30 THz vibrational frequency of the C-F stretch mode of CHFClBr molecules using saturated absorption spectroscopy [14,15]. This reached a precision of 8 Hz, a great achievement but still a factor of 3000 larger than the ~2 mHz predicted frequency difference between the left- and right-handed enantiomers [16]. To improve on this, we are building a more sensitive instrument and are learning to work with molecules where the energy difference is expected to be far larger.

The paper is organized as follows. Section 2) discusses the new promising oxorhenium organo-metallic species we are learning to work with. Section 3) gives an overview of the new experiment. Section 4) and 5) describe the development of buffer-gas sources of slow, cold polyatomic molecules, and of ultra-stable quantum cascade lasers calibrated on primary atomic frequency standards, respectively.

**2) Molecules considered**

The PV energy difference scales strongly with nuclear charge [17], so we consider molecules with heavy atoms near the chiral centre. We have already worked with theoretical and experimental chemists to find the best chiral species for measuring the PV energy difference. This led to the successful synthesis of solid oxorhenium organo-metallic compounds that have vibrational transitions around 30 THz and PV frequency shifts as large as ~1 Hz (see [5,18-21] and references therein), up to 1000 times larger than in CHFClBr. Several other tracks are currently being followed and other molecules are being prepared. Our collaborators have recently found a uranium compound (N≡UHFI) with a record ~20 Hz predicted PV frequency shift [22]. Although synthesizing and isolating such compounds has not been demonstrated so far, this could be possible in the future.

To gain insight in the apparatus and know-how required for performing precise spectroscopic measurements on such complex species, we have conducted high-resolution mid-infrared spectroscopy of an achiral precursor methyltrioxorhenium (MTO, $CH_3ReO_3$) in both room temperature cells and cold supersonic beams [5,18,20,23,24]. Building on these first results, we are designing and constructing a state-of-the-art instrument for precise vibrational spectroscopy of chiral molecules.

3) **Overview of the apparatus**

Figure 1 illustrates the set-up of this new experiment. A slow, cold molecular beam of the rhenium species of interest is made by producing the molecules inside a buffer-gas cell cooled to a few kelvins, and then extracting the molecules into a beam. Such buffer-gas beams of solid-state molecules formed in a cryogenic cell, one of the latest molecular beam source technologies, exhibit both low velocity and some of the highest beam fluxes to date [25], making them very attractive for precise spectroscopic measurements. The molecules then pass through a Ramsey interferometer sensitive enough to measure tiny changes in the vibrational frequency of the molecules [26]. To reach the required frequency, which is in the mid-infrared region of the spectrum, we use quantum cascade lasers (QCLs). To achieve the required frequency stability, we stabilise the laser to a frequency comb, which is itself locked to an ultra-stable 1.5 µm laser signal, ultimately referenced to primary frequency standards, potentially a Cs fountain clock which realizes the International System of Unit standard of time [27,28]. QCLs offer broad and continuous tuning, and available wavelengths cover the entire mid-infrared region allowing the study of a considerable number of candidate species and thus providing invaluable flexibility. Molecules must then be detected with high sensitivity. We are currently investigating detectors based on cavity-enhanced schemes recording mid-infrared absorption or the microwave field molecules emit when they rotate. To measure the PV vibrational frequency shift, the vibrational frequency of each enantiomer must be measured in this new apparatus. Because this is a differential measurement between the two enantiomers, most systematic frequency shifts will cancel out. We project a measurement precision below 0.1 Hz [5], an improvement of at least two orders of magnitude compared to the CHFClBr experiment.

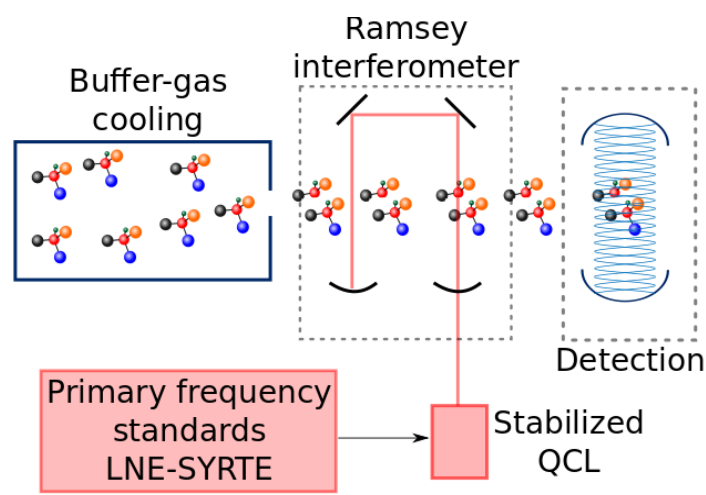

**Figure 1:** Illustration of the proposed experiment to measure parity-violating rovibrational frequency differences between enantiomers of cold chiral molecules. QCL: quantum cascade laser.

### 4) Buffer gas cooling

We have already made much progress in the development of methods to cool the complex and heavy rhenium species of interest for PV measurements to a few kelvins by collisions with a buffer gas of 4 K helium inside a copper cell mounted on the cold stage of a cryo-cooler. As illustrated in Figure 2, we have demonstrated cryogenic buffer-gas cooling of the first organo-metallic species, MTO [24,29]. The molecules are produced with a rotational temperature of approximately 6 K by laser ablation of a MTO pellet inside the buffer-gas cell. This development extends the technique of buffer-gas cooling to a new class of molecules. We have also learned how to bring such species into the gas-phase by laser ablation, a method previously used only for diatomic molecules. Many polyatomic molecules have little vapour pressure at easily accessible temperatures, so the ability to introduce them by laser ablation is important. Our work shows that rhenium species of interest for PV measurements survive the ablation process and that the translational and rotational degrees of freedom cool efficiently through collisions with helium, which is very promising for the production of buffer-gas-cooled molecular beams. We have also demonstrated the first precise spectroscopic measurements of buffer-gas-cooled molecules in the mid-infrared molecular fingerprint region around 10 μm, obtaining rotational and hyperfine-resolved absorption spectra (see Figure 2) and allowing the determination of the nuclear quadrupole coupling of the excited vibrational state, which is unprecedented for such a complex molecule. We have also demonstrated cryogenic buffer-gas cooling of another polyatomic molecule, 1,3,5-trioxane (($CH_2O$)$_3$, a cyclic trimer of formaldehyde). Although a solid at room temperature, it has enough volatility to be injected into the buffer-gas cell through a room temperature tube. We have recorded saturated absorption spectra of trioxane which constitute the first sub-Doppler spectroscopic measurements of buffer-gas-cooled molecules in the fingerprint region and the first sub-Doppler spectra of polyatomic molecules inside a buffer gas cell.

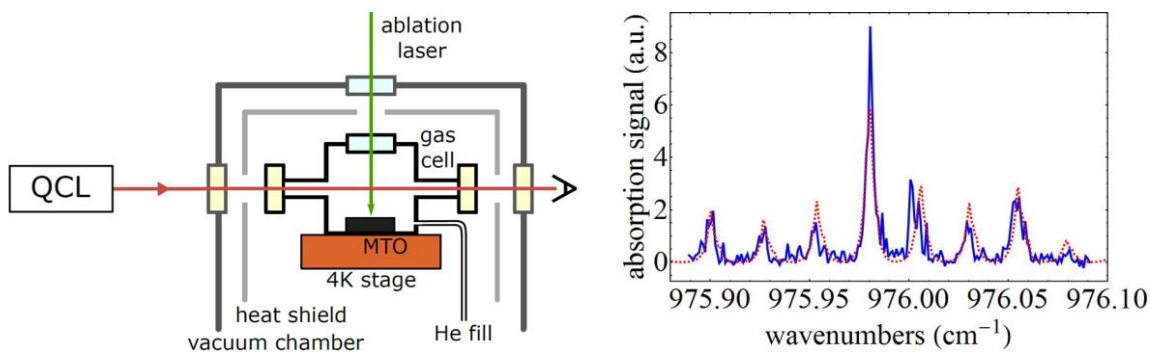

**Figure 2:** *Left:* Experimental setup for cryogenic buffer-gas cooling of methyltrioxorhenium (MTO). A copper cell is filled with ~$10^{-2}$ mbar of helium and is cooled to 6 K using a cryocooler. The cell is surrounded by aluminium radiation shields at a temperature of about 40 K, and is housed in a vacuum chamber. A target containing MTO is ablated with pulses from a Nd:YAG laser at 1064 nm. Light from a free-running quantum cascade laser (QCL) emitting at ~976 cm$^{-1}$ is sent through the cell to record absorption spectra. *Right:* Buffer-gas-cooled spectrum showing the rotational contour of the *Q* branch of the antisymmetric Re=O stretching mode of the $^{187}$Re isotopologue. The dashed (red) curve is a fit to the data from which a rotational temperature of 6 K ± 3 K is inferred [29].

Challenges remaining in the source development are the construction of a versatile buffer-gas source for beam production. Our work has focussed on developing methods of laser ablation and capillary loading. All the measurements were performed inside the cell. The next step will be to make a molecular beam, and optimize the cell configuration to create a slow, high-flux beam of cold complex molecules. Given that we do not yet know which chiral molecule we will use for our measurement, it is beneficial to design a versatile cell that can be used with a range of loading mechanisms. Capillary loading is well-suited to molecules with a high volatility, whereas low volatility species are better introduced *via* an oven or through laser ablation. Figure 3 shows different loading mechanisms that we aim to investigate.

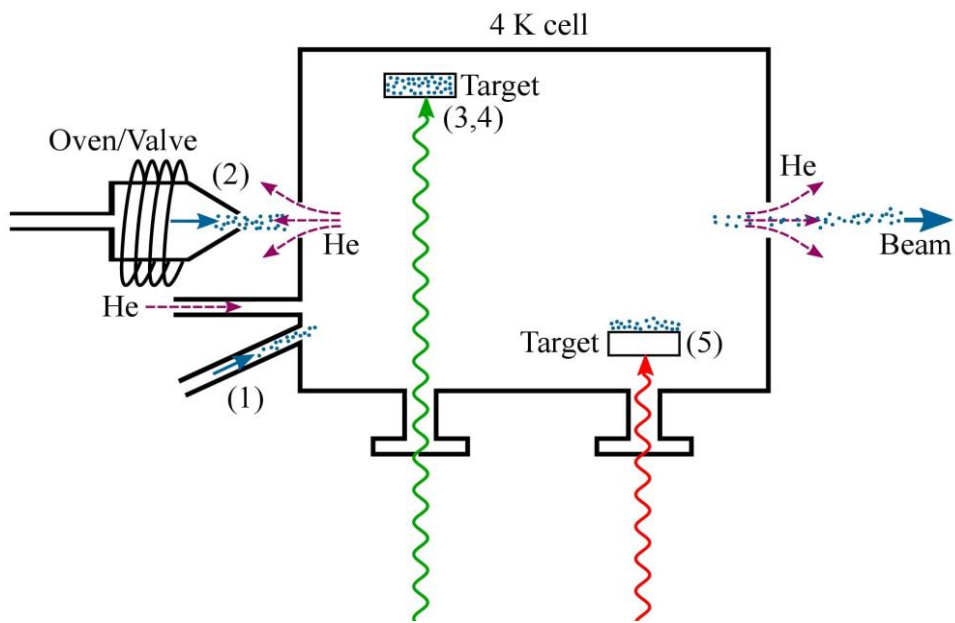

**Figure 3:** Schematic diagram of a buffer-gas cell, showing different molecule-loading mechanisms: (1) capillary loading, (2) loading from a valve or oven, (3) laser ablation, (4) matrix-assisted desorption, (5) laser-induced acoustic desorption. Blue: molecules of interest; purple: helium; green: laser for ablation or desorption of a solid target containing the molecules of interest; red: laser for laser-induced acoustic desorption of a solid target containing the molecules of interest.

The development of the molecular beam will require sensitive detection. We will build an optical cavity around the beam-line and perform cavity-enhanced absorption spectroscopy in the mid-infrared. Combining this with wavelength modulation spectroscopy will allow weak absorption signals to be recorded with a large signal-to-noise ratio, allowing great sensitivity for the optimization of the beam flux.

5) **Laser stabilization for precise vibrational spectroscopy**

Regarding the development of laser techniques for our new instrument, we have been able to phase-lock a 10 µm QCL to the secondary frequency standard of this spectral region, a $CO_2$ laser stabilised on a saturated absorption line of $OsO_4$ [23,30]. The excellent spectral features of the ultra-stable $CO_2$ laser were transferred to the QCL, resulting in a line-width on the order of 10 Hz, about two orders of magnitude narrower than the previously published narrowest QCL. More recently, we have developed a widely tuneable frequency-stabilised QCL with direct traceability to primary frequency standards. Our method allows us to lock any mid-infrared laser to a frequency comb stabilised to a near-infrared reference, given by the optical phase of an ultra-stable 1.55 µm laser located at LNE-SYRTE (the French national metrology institute). This signal is monitored against atomic frequency standards and transferred via a 43-km long optical fibre cable with correction of the propagation-induced phase noise. The stability of the reference is transferred to the mid-infrared source, which therefore exhibits a relative frequency stability lower than $2 \times 10^{-15}$ between 1 and 100 s. Moreover, thanks to the traceability to the primary standards of LNE-SYRTE, its absolute frequency is known with an uncertainty below $10^{-14}$ (potentially as low as the accuracy of the Cs fountain clock, $3 \times 10^{-16}$). This stabilisation method was demonstrated with a $CO_2$ laser [31,32] and a QCL [27,28]. Stabilising the laser this way frees us from having to lock the QCL to any particular reference (either another mid-infrared laser or a molecular transition), which would constrain the laser's operating frequency range. It also results in a ~0.1-Hz line-width. To our knowledge, this is by far the narrowest and most accurate QCL reported. The setup allows the QCL to be scanned over a few hundred megahertz while maintaining the highest stabilities and accuracies by scanning the near-infrared reference using an electro-optic modulator [33]. To demonstrate the QCL's potential for state-of-the-art precise spectroscopic measurements, saturated absorption spectra of $OsO_4$ as narrow as 25 kHz were recorded using a Fabry-Perot cavity, allowing central frequencies to be determined with an uncertainty of a few tens of hertz [27,28]. As illustrated in Figure 4, we have also conducted saturated absorption spectroscopy of several methanol lines in a multi-pass cell, and have determined central frequencies with an uncertainty of a few kilohertz, an improvement of 3 to 4 orders of magnitude over previous measurements.

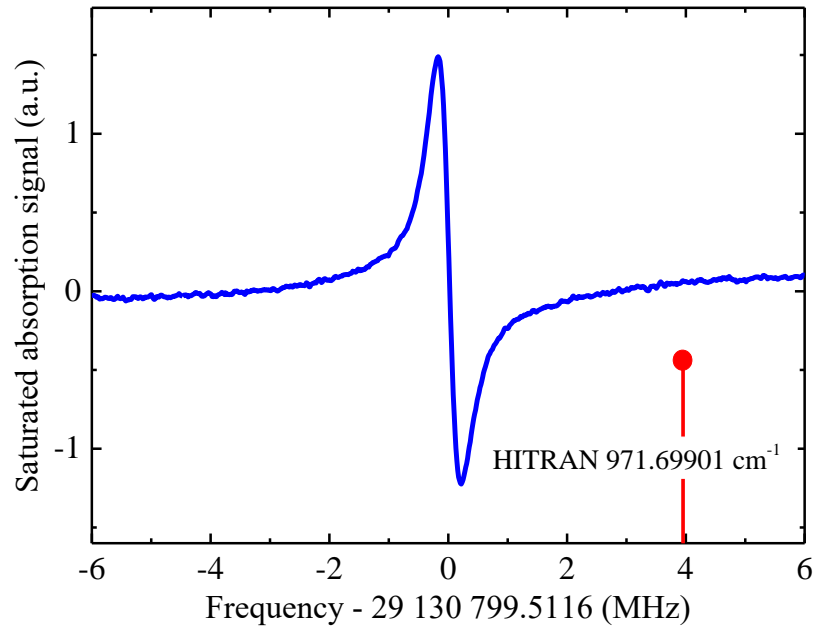

**Figure 4:** Saturated absorption spectrum of the P(*A*,co,0,0$^+$,33) rovibrational line of methanol, belonging to the *P* branch of the $\nu_8$ C-O stretch vibrational mode, recorded in an astigmatic Herriott multipass cell (Aerodyne Research, model AMAC-36, 20 cm base path, 182 paths) using frequency modulation and first-harmonic detection. Light from a stabilised QCL is split into a pump beam and a probe beam of power incident to the cell of around 1.5 mW and 0.8 mW respectively. Experimental conditions: pressure: 1 Pa; modulation frequency: 20 kHz; frequency modulation excursion; 50 kHz; frequency step: ~15 kHz; average of 5 pairs of scans of opposite frequency sweep direction; total integration time per point: 1 s. The P(*A*,co,0,0$^+$,33) line frequency position reported in the HITRAN database [34] is also shown as a red stick.

We have so far demonstrated ultra-stable QCLs in a limited spectral window around 10 μm, but the resonances of many promising molecules are not at 10 μm. Challenges in the laser system development are the construction of fully operational spectrometers based on QCLs stabilised to the comb anywhere in the mid-infrared while taking full advantage of the wide tuneability offered by QCLs. To provide a high level of flexibility, the spectrometer should be designed in such a way that QCLs can be easily interchanged, to allow us to stabilize QCLs at any frequency.

6) **Summary**

We have described the progress towards our new experiment to measure signatures of parity violation in chiral molecules. We have discussed recent developments made in the production of cryogenically cold complex molecules, and in laser stabilization of quantum cascade lasers. Together, the advances reported here demonstrate some of the methods needed for a measurement of parity violation in chiral molecules. Beyond PV, they also open up new possibilities for measuring and controlling complex molecules and for using them in a variety of precise measurements.

**Acknowledgments:**

The authors thank P. Asselin, R. Bast, C. Chardonnet, J. Crassous, C. Daussy, L. Guy, E. A. Hinds, T. R. Huet, T. Saue, T. J. Sears, P. Soulard and J. Tandy for fruitful discussions. In France, this work was supported by ANR (under grants no. ANR 2010 BLAN 724 3, no. ANR-12-ASTR-0028-03, no. ANR-15-CE30-0005-01 no. ANR-17-ERC2-0036-01, and through Labex First-TF ANR 10 LABX 48 01), through EMPIR projects 15SIB05 "OFTEN" and 15SIB03 "OC18" (this project has received funding from the EMPIR programme co-financed by the Participating States and from the European Union's Horizon 2020 research and innovation programme) and Région Île-de-France (DIM Nano-K), CNRS, Université Paris 13 and AS GRAM. In the UK, the work was supported by EPSRC under grant

EP/I012044/1. The work was made possible through the International Exchanges Programme run by the Royal Society and CNRS. D.B.A. Tran is supported by the Ministry of Education and Training, Vietnam (Program 911).**References:**

[1] Rein D. W., J. Mol. Evol., 4, 15 (1974).

[2] Quack M., Stohner J., Willeke M., Annu. Rev. Phys. Chem., 59, 741 (2008).

[3] MacDermott A. J., Fu T., Hyde G. O., Nakatsuka R., Coleman A. P., Orig. Life Evol. Biosphere, 39, 407 (2009).

[4] Barkov L. M., Zolotorev M. S., Phys. Lett., 85B, 308 (1979).

[5] Darquié B., Stoeffler C., Shelkovnikov A., Daussy C., Amy-Klein A., Chardonnet C., Zrig S., Guy L., Crassous J., Soulard P., Asselin P., Huet T. R., Schwerdtfeger P., Bast R., Saue T., Chirality, 22, 870 - 884 (2010).

[6] Gonzalo I., Bargueno P., de Tudela R. P., Miret-Artes S., Chem. Phys. Lett., 489, 127 (2010).

[7] Medcraft C., Wolf R., Schnell M., Angew. Chem., Int. Ed. Engl., 53, 11656 (2014).

[8] S. Eibenberger, J. Doyle, D. Patterson, Phys. Rev. Lett., **118**, 123002 (2017).

[9] Eills J., Blanchard J. W., Bougas L., Kozlov M. G., Pines A., Budker D., Phys. Rev. A, 96, 042119 (2017).

[10] Quack M., Chem. Phys. Lett., 132, 147 (1986)

[11] Dietiker P., Miloglyadov E., Quack M., Schneider A., Seyfang G., J. Chem. Phys., 143, 244305 (2015).

[12] Fábri C., Horný L., Quack M., ChemPhysChem, 16, 3584 (2015).

[13] Letokhov V. S., Phys. Lett. A, 53, 275 (1975).

**Addresses:**

[1]Laboratoire de Physique des Lasers, CNRS, Université Paris 13, Sorbonne Paris Cité, 99 avenue Jean-Baptiste Clément, 93430 Villetaneuse, France

[2]LNE-SYRTE, Observatoire de Paris, PSL Research University, CNRS, Sorbonne Universités, 61 Avenue de l'Observatoire, 75014 Paris, France

[3]Centre for Cold Matter, Blackett Laboratory, Imperial College London, Prince Consort Road, London SW7 2AZ, United Kingdom

[†]permanent address: Faculty of Physics, Ho Chi Minh City University of Education, Ho Chi Minh City, Vietnam

[§]present address: ONERA, The French Aerospace Lab., Centre de la Hunière, BP 80100, Palaiseau 91123, France

[#]present address: Laboratoire Kastler Brossel, Sorbonne Université, CNRS, ENS-PSL Research University, Collège de France, 4 place Jussieu, 75252 Paris, France

[£]permanent address: Institute of Laser Physics of the Siberian Branch of the Russian Academy of Sciences, Pr. Lavrentyeva 13/3, Novosibirsk, 630090 Russia

[$]present address: Real Instituto y Observatorio de la Armada, San Fernando, Spain

[!]present address: National Physical Laboratory, NPL, Hampton Road, Teddington TW11 0LW, United Kingdom

E-mail: benoit.darquie@univ-paris13.fr